\documentclass[10pt,aps,prd,floatfix,notitlepage,showpacs,nofootinbib,superscriptaddress,showkeys]{revtex4-1}
\usepackage{amssymb,amsmath}
\usepackage[colorlinks=true, colorlinks, linkcolor=blue, citecolor=magenta]{hyperref}
\usepackage{graphicx}

\begin{document}
\title{Linearized Field Equations and Extra Force in $f(R,{\bf T}^{(n)})$ Extended  Gravity}

\author{Habib Abedi}
\email{h.abedi@ut.ac.ir}
\affiliation{Department of Physics, University of Tehran, North Kargar Avenue, 14399-55961 Tehran, Iran.}

\author{Francesco Bajardi}
\email{francesco.bajardi@unina.it} 
\affiliation{Scuola Superiore Meridionale, Largo San Marcellino 10, 80138 Napoli, Italy}
\affiliation{INFN Sezione  di Napoli, Compl. Univ. di Monte S. Angelo, Edificio G, Via Cinthia, I-80126, Napoli, Italy}

\author{Salvatore Capozziello}
\email{capozziello@na.infn.it}
\affiliation{Scuola Superiore Meridionale, Largo San Marcellino 10, 80138 Napoli, Italy}
\affiliation{Dipartimento di Fisica "Ettore Pancini", Universit\`{a} degli Studi di Napoli
"Federico II", Compl. Univ. di Monte S. Angelo, Edificio G, Via Cinthia, I-80126, Napoli, Italy}
\affiliation{INFN Sezione  di Napoli, Compl. Univ. di Monte S. Angelo, Edificio G, Via Cinthia, I-80126, Napoli, Italy}
\affiliation{Laboratory  for Theoretical Cosmology, International Centre of Gravity and Cosmos,
Tomsk State University of Control Systems and Radioelectronics (TUSUR),
634050 Tomsk, Russia.}

\begin{abstract}
We consider an extended  theory of gravity with Lagrangian $\mathcal{L} = f(R,{\bf T}^{(n)})$, with ${\bf T}^{(n)}$ being a $2n$-th order invariant made of contractions of the energy-momentum tensor. When $n=1$ this theory reduces to $f(R,T)$ gravity, where $T$ accounts for the trace of the energy-momentum tensor. We study the gravitational wave polarization modes, from which it results that when the matter Lagrangian contains dynamical scalar fields minimally coupled to the geometry, further polarization modes arise with respect to General Relativity. Finally we show that the motion for test particles is non-geodesic and we explicitly obtain the extra-force. 

\end{abstract}

	\date{\today}

\date{\today}

\maketitle
\section{Introduction}
Throughout the years from its formulation, General Relativity (GR) gained several successes at any scale of energy, as it was the first theory describing the gravitational interaction by means of the space-time geometry. Recently, the Gravitational Waves (GWs) detection~\cite{LIGOScientific:2016aoc, LIGOScientific:2016vlm, LIGOScientific:2016wkq} and the Black Holes (BHs) observation~\cite{EventHorizonTelescope:2019dse} further corroborated the validity of the description of gravity in terms of curvature. In spite of this wide success, during more than one hundred years, experiments and  observations provided some incompatibilities with the theory, to the point of calling its validity into question~\cite{Joyce:2014kja, Capozziello:2011et}. As a matter of facts, in the large-scale regime the today observed accelerating expansion of the Universe cannot be predicted without introducing the so called dark energy~\cite{Peebles:2002gy, Padmanabhan:2002ji}, as well as the Galaxies structure suffers several shortcomings without introducing dark matter~\cite{ParticleDataGroup:2012pjm, Bosma:1981zz}. These are only two examples of the problems suffered by GR; for a review on the topic see \emph{e.g.}~\cite{ Faraoni:2010pgm, Nojiri:2006ri, Nojiri:2010wj}. At small scales, GR formalism cannot be merged with that of Quantum Mechanics, due to the lack of a full consistent theory of Quantum Gravity. Several attempts have been pursued in this direction (\emph{e.g.} non-local theories~\cite{Arkani-Hamed:2002ukf, Modesto:2013jea, Capozziello:2022lic}, String Theory~\cite{Polchinski:1998rq}, Loop Quantum Gravity~\cite{Ashtekar:2011ni, Rovelli:1997yv}, \emph{etc.}), but all of them lack at least one necessary condition among full renormalizability, unitarity and physical predictability. For all these reasons, several different gravitational theories have been developed, with the aim of addressing issues provided by GR. Most of them relaxes some assumptions of GR, such as Lorentz invariance~\cite{Kiritsis:2009sh, Colladay:1998fq}, Equivalence Principle~\cite{Arcos:2004tzt, Krssak:2018ywd, DeAndrade:2000sf}, or second-order field equations~\cite{Capozziello:2011et, DeFelice:2010aj}. The latter category considers extensions of the Hilbert-Einstein action, containing \emph{e.g.} functions of the scalar curvature~\cite{Sotiriou:2008rp, Clifton:2011jh}, higher-order curvature invariants~\cite{Amendola:1993bg, Bajardi:2020osh}, non-minimal coupling between geometry and dynamical scalar fields~\cite{Bajardi:2020xfj, Halliwell:1986ja, Uzan:1999ch, Rubano:2001su}, \emph{etc.} In all cases, the variation of the extended action with respect to the metric tensor provides extra terms in the higher-order field equations, which can be recast as an effective energy-momentum tensor of the gravitational field. In this way, dark energy and dark matter can be mimicked by geometric contributions~\cite{Copeland:2006wr, Capozziello:2002rd, Mantica} and additional potentials arise in the weak field limit \cite{Capozziello:2021goa}. Moreover, from the one-loop effective action of GR, it turns out that renormalizability requires higher-order corrections to allow the UV divergences canceling out. Higher-order corrections lead to a positive mass dimension of the coupling constant when the action is expanded around a Gaussian fixed point~\cite{Pawlowski:2018ixd, Niedermaier:2006wt, Bajardi:2021lwp}.

An interesting class of alternatives to GR comes considering the coupling between geometric terms and matter sectors, such as the model with Lagrangian
\begin{equation}
\mathcal{L} = f_1(R)+ \left[ 1+ \lambda \, f_2(R) \right] {\cal L}_{\rm m},
\end{equation}
where $\lambda$ is a coupling constant providing the strength of the interaction between $f_2(R)$ and the matter Lagrangian ${\cal L}_{\rm m}$. This theory has been studied \emph{e.g.} in Refs.~\cite{Bertolami:2007gv, Harko:2010zi}. Such coupling leads to a non-conservation of the matter energy-momentum tensor and, consequently, yields an extra force orthogonal to the four-velocity. In Ref.~\cite{Harko:2008qz} this theory was further extended to $f_1(R)+G({\cal L}_{\rm m})\, f_2(R)$, while in Ref.~\cite{Harko:2010mv} the authors consider the general Lagrangian containing the function $f(R,{\cal L}_{\rm m})$. In Ref.~\cite{Nagpal:2019vre} another modification including an arbitrary function $f(R,T)$, with $T$ being the trace of the energy-momentum tensor, is considered by the authors. 

In this context, Harko \emph{et al.}~\cite{Harko:2011kv} proposed the function $f(R,T^\phi)$, where $T^\phi$ is the trace of the scalar field energy-momentum tensor. Katirci and Kavuk, in Ref.~\cite{Katirci:2013okf}, introduced an extension of this model, by considering the function $f(R,T^2)$, with $T^2:= T_{\mu\nu}T^{\mu\nu}$. Finally, in Ref.~\cite{Haghani:2013oma}, Haghani \emph{et al.} studied  the model $f(R,T,R_{\mu\nu}T^{\mu\nu})$.

In this paper,  we consider an extended action containing a function of the scalar curvature and of the higher-order invariant ${\bf T}^{(n)}$, defined by means of the energy-momentum tensor $T_{\mu \nu}$ as:
\begin{align}
	{{\bf T}^{\mu}_{\phantom{\mu}\nu}}^{(n)}:=& T^{\mu}_{\phantom{\mu}\alpha_1} \, T^{\alpha_1}_{\phantom{\alpha_1}\alpha_2} \, \cdots \, T^{\alpha_{n-1}}_{\phantom{\alpha_{n-1}}\nu},
	\\
	{\bf T}^{(n)}:=&{{\bf T}^{\mu}_{\phantom{\mu}\mu}}^{(n)}.
\end{align}
This theory is clearly a generalization of the above mentioned $f(R,T)$ or $f(R,T^2)$ gravity.
The presence of the energy-momentum tensor in the gravitational action can be understood as generated by exotic fluids or quantum effects~\cite{Adams:1990pn, Cotsakis:2006zn}. This type of models have been largely studied \emph{e.g.} in Refs.~\cite{Haghani:2013oma, Harko:2021tav, Sharif:2012zzd, Houndjo:2011tu, Yousaf:2016lls, Shabani:2013djy, Moraes:2015uxq}, where astrophysical and cosmological applications are considered to investigate the large-scale structures. Here we consider the $f(R, {\bf T}^{(n)})$ function to find the linearized field equations and thus the GWs equation. Then we study the motion of a test particle, showing that an extra force, orthogonal to the velocity, occurs. Nowadays, GWs represent a fundamental testbed for the validity of any gravitational theory, and the research for possible new modes accounts for an active topic within the context of modified theories of gravity. For instance, in Ref.~\cite{Liang:2017ahj}, GWs are studied in the context of $f(R)$ gravity, where it turns out that three additional polarization modes occur.
In Ref.~\cite{Abedi:2017jqx} modified teleparallel gravity and its extensions are considered, with the result that modified $f(\mathcal{T})$ theory (with $\mathcal{T}$ being the torsion scalar) carries exactly the same mode as standard Einstein gravity. In~\cite{Capozziello:2006ra, Liu:2019cxm} the same prescription is applied to scalar-tensor models, while in~\cite{Holscher:2018jhm, Capriolo1, Capriolo2} higher-order terms are introduced in the gravitational action. For a review on GWs in modified theories of gravity see \cite{Capozziello:2019klx}. 
%
 %
 The number of GWs  polarization were also studied for $f(R,T)$ and $f(R,T^\phi)$ models in Ref.~\cite{Alves:2016iks}, where $T$ and $T^\phi$ are the trace of the energy-momentum tensor of standard matter and of a scalar field, respectively. 

This paper is organized as follows: in Sec.~\ref{Sec:FE}, we introduce the $f(R, {\bf T}^{(n)})$ model and find out the gravitational field equations in the metric formalism, also showing that the matter energy-momentum tensor is not conserved. 
In Sec.~\ref{Sec:LinearizedEq}, we focus on a matter Lagrangian depending on the dynamical scalar field $\phi$ and get the explicit form of the energy-momentum tensor. We then adopt the first-order approximation of the field equations to obtain the wave equation. As we will show, the non-vanishing divergence of the matter energy-momentum implies a non-geodesic motion for the test particles. In Sec.~\ref{Sec:EF} we obtain the extra force for the $f(R, {\bf T}^{(n)})$ model and finally, in Sec.~\ref{Sec:concl}, we conclude the work with discussions and future perspectives.

\section{Field equations in $f(R, {\bf T}^{(n)})$ Gravity}
\label{Sec:FE}
Let us derive the field equations for the model $f(R, {\bf T}^{(n)})$, with the aim to adopt the first-order approximation and study the related GWs modes. To this purpose, we consider a general function $f(R, {\bf T}^{(n)})$ depending on the integer $n$. Let us start from the following higher-order gravitational action
\begin{align} \label{action}
	{\cal S} = \int \, {\rm d}^4x \; \sqrt{-g} \left[\frac{1}{2\kappa^2} f\left(R,{\bf T}^{(n)}\right) + {\cal L}_{\rm m} \right], 
\end{align}
where $\kappa^2 \equiv 8 \pi G_N$, with $G_N$ being the Newton constant. Here $ f\left(R,{\bf T}^{(n)}\right) $ is a function of the scalar curvature $R$ and of the higher-order invariant ${\bf T}^{(n)}$, while $g$ denotes the determinant of the metric tensor. Assuming the matter Lagrangian ${\cal L}_{\rm m}$ to be only dependent on the metric tensor and not on its derivatives, the energy-momentum tensor $T^{\mu \nu}$ can be defined as:
\begin{align}
	T^{\mu\nu} = -\frac{2}{\sqrt{-g}} \frac{\delta\left(\sqrt{-g} {\cal L}_{\rm m}\right)}{\delta g^{\mu\nu}} = g^{\mu\nu} {\cal L}_{\rm m} -2 \frac{\partial {\cal L}_{\rm m}}{\partial g^{\mu\nu}}.
	\label{EMtens}
\end{align}
Starting from Eq. \eqref{EMtens}, we define the $2n$-th order invariant ${\bf T}^{(n)}$ by means of the contraction of the energy-momentum tensor, as:
\begin{align}
	{{\bf T}^{\mu}_{\phantom{\mu}\nu}}^{(n)}:=& T^{\mu}_{\phantom{\mu}\alpha_1} \, T^{\alpha_1}_{\phantom{\alpha_1}\alpha_2} \, \cdots \, T^{\alpha_{n-1}}_{\phantom{\alpha_{n-1}}\nu},
	\label{Tnmunu}
	\\
	{\bf T}^{(n)}:=&{{\bf T}^{\mu}_{\phantom{\mu}\mu}}^{(n)}. \label{Tn}
\end{align}
Notice that for $n=1$ we have $T^{\mu}_{\phantom{\mu}\nu}= {{\bf T}^{\mu}_{\phantom{\mu}\nu}}^{(1)} $, while for $n=2$ we recover the so called ``squared gravity''. The variation of the action~\eqref{action} with respect to the metric gives
\begin{align}
	\delta {\cal S}_{\rm g} =\frac{1}{2\kappa^2} \int \, {\rm d}^4x \; \sqrt{-g} \left[  - \frac{1}{2} g_{\mu\nu} f \, \delta g^{\mu\nu} + f_R  \, \delta R+f_{{\bf T}^{(n)}} \, \delta {\bf T}^{(n)} \right],
\end{align}
where we defined $f_{{\bf T}^{(n)}}:=\partial f / \partial {\bf T}^{(n)}$, $f_R:= \partial f / \partial R$. Discarding total derivatives and using the well known Ricci scalar variation 
\begin{align}
\delta R = R_{\mu\nu} \, \delta g^{\mu\nu} + g_{\mu\nu}\, \square \, \delta g^{\mu\nu}- \nabla_\mu \nabla_\nu \, \delta g^{\mu\nu},
\end{align}
the field equations can be written as:
\begin{align}
f_R R_{\mu\nu}-\frac{1}{2}g_{\mu\nu} f+ \left(g_{\mu \nu}\square - \nabla_\mu \nabla_\nu\right) f_R =& \kappa^2 T_{\mu \nu} - f_{{\bf T^{(n)}}} \Theta_{\mu \nu}^{(n)}, \label{FE}
\end{align}
so that the trace equation reads
\begin{align}
	f_R R -2f + 3 \square f_R = \kappa^2 T -f_{\bf T^{(n)}} \Theta^{(n)},
\end{align}
with the definitions $\Theta_{\mu \nu }^{(n)}:=  \delta {\bf T}^{(n)}/\delta g^{\mu \nu}$  and $\square = \nabla^\mu \nabla_\mu$.
If $f(R,{\bf T}^{(n)}) = f(R)$ we recover the $f(R)$-gravity field equations. 
The above equation suggests that it is possible to define an effective energy-momentum tensor, comprehending both matter and geometric contributions, namely
\begin{align}
	T_{\mu\nu}^{\rm (eff)} = T_{\mu\nu} - \kappa^{-2} f_{\bf T^{(n)}} \Theta_{\mu \nu}^{(n)}.
\end{align}

The quantity $\Theta_{\mu \nu }^{(n)}$ can be explicitly computed by considering that ${\cal L}_{\rm m}$ is independent of the metric tensor derivatives, so that its variation yields 
\begin{align}
\delta {\cal L}_{\rm m} = \left(\frac{1}{2} g_{\mu\nu} {\cal L}_{\rm m}  -\frac{1}{2} T_{\mu\nu} \right) \delta g^{\mu\nu},
\end{align}
from which the relation below automatically follows:
\begin{align}
	\frac{\delta T^\alpha_{\phantom{\alpha}\beta}}{\delta g^{\mu\nu}} = T_{\nu\beta} \, \delta^\alpha_\mu +g^{\alpha \sigma} \left[-\frac{1}{2} {\cal L}_{\rm m} \left(g_{\sigma \mu} g_{\beta \nu}-g_{\sigma \nu}g_{\beta \mu}\right)+g_{\sigma\beta}\left(\frac{1}{2} g_{\mu\nu}{\cal L}_{\rm m}-\frac{1}{2}T_{\mu\nu}\right)-2\frac{\partial^2 {\cal L}_{\rm m}}{\partial g^{\mu\nu} \, \partial g^{\sigma\beta}}\right].
	\label{variationTmunu}
\end{align}
Therefore, replacing Eq. \eqref{variationTmunu} into the definition of the rank-2 tensor $\Theta_{\mu \nu }^{(n)}$, we finally get
\begin{align}
	\Theta_{\mu \nu }^{(n)}=& n {\bf T}_{\mu\nu}^{(n)}
	-n {\cal L}_{\rm m}  {\bf T}_{\mu\nu}^{(n-1)}
	+ n \left(\frac{1}{2} g_{\mu\nu} {\cal L}_m - T_{\mu\nu}\right) {\bf T}^{(n-1)}
	-2 n g^{\alpha\gamma} \frac{\partial^2 {\cal L}_{\rm m}}{\partial g^{\mu \nu} \, \partial g^{\beta \gamma}} \, {{\bf T}^\beta_{\phantom{\beta}\alpha}}^{(n-1)} , \label{thetamunu}
	\\
	\Theta^{(n)} =&g^{\mu\nu} \Theta^{(n)}_{\mu\nu}\nonumber\\=& n {\bf T}^{(n)} + n {\bf T}^{(n-1)} \left({\cal L}_m - T \right) -2 n g^{\mu \nu} g^{\alpha \gamma} \frac{\partial^2 {\cal L}_{\rm m}}{\partial g^{\mu \nu} \, \partial g^{\beta \gamma}}  \, {{\bf T}^\beta_{\phantom{\beta}\alpha}}^{(n-1)}.
	\label{tracetheta}
\end{align}
Squared gravity is obtained for $n=2$  in  Eq. \eqref{thetamunu}, which yields:
\begin{align}
	{\bf T}^{(2)}=& T^{\alpha_1}_{\phantom{\alpha_1}\alpha_2} \, T^{\alpha_2}_{\phantom{\alpha_1}\alpha_1},
	\\
	\Theta^{(2)}_{\mu\nu} =& -2 {\cal L}_{\rm m}\left(T_{\mu\nu}-\frac{1}{2} g_{\mu\nu}T\right) - T T_{\mu\nu} +2T^{\phantom{\nu}\alpha}_\nu \, T_{\alpha\nu} - 4 T^{\alpha \beta} \frac{\partial^2 {\cal L}_{\rm m}}{\partial g^{\mu\nu} \, \partial g^{\alpha \beta}}.
\end{align}
The covariant derivative of Eq.~\eqref{FE}, along with the Bianchi identities $\nabla^\mu G_{\mu\nu}=0$ and
\begin{align}
    \left(\nabla_\nu  \, \square -\square \, \nabla_\nu \right) f_R =  -R_{\mu\nu} \, \nabla^\mu f_R
\end{align}
permits to derive the non-conservation of the energy-momentum tensor, that is:
\begin{align}\label{cons}
	\kappa^2 \nabla_\mu T^\mu_\nu = \nabla^\mu \left(f_{{\bf T}^{(n)}} \Theta^{(n)}_{\mu\nu}\right) - \frac{1}{2} f_{{\bf T}^{(n)}} \nabla_\nu {\bf T}^{(n)},
\end{align}
where we have used the relation
$\nabla_{[\mu}\nabla_{\nu]} V_\alpha = R^\sigma_{\phantom{\sigma}\alpha \nu\mu} V_\sigma$ for an arbitrary vector field $V_\mu$. 
Eq.~\eqref{cons} is interesting, as the dynamical analysis may result in some new effects. 
For $f(R,T^\phi)$, see Ref.~\cite{Singh:2018tlm}. Here the authors constrained the field equations by a specific \emph{ansatz} on the shape of  starting function. For $f(R,{\bf T}^{(n)}) = f(R)$, the standard energy-momentum tensor conservation $\nabla^\mu T_{\mu\nu} =0$ is recovered, as we expect.

The field equations can be recast in a more suitable form, by assuming an \emph{ansatz} for the function $f(R,{\bf T}^{(n)})$. Specifically, we can consider the case $f(R,{\bf T}^{(n)})= R+ F({\bf T}^{(n)})$, where GR is safely recovered as soon as $F({\bf T}^{(n)})$ vanishes. In this way, the latter can be intended as an effective cosmological constant provided by geometry. By this choice, Eq. \eqref{variationTmunu} reduces to:
\begin{align}
G_{\mu\nu} =& \kappa^2 T_{\mu\nu} +\frac{1}{2} g_{\mu\nu} F - F_{{\bf T}^{(n)}} \Theta_{\mu \nu}^{(n)},\label{FEq}
\end{align}
with trace equation of the form
\begin{align}
-R =& \kappa^2 T^\alpha_{\phantom{\alpha}\alpha} +2 F - F_{{\bf T}^{(n)}} \Theta^{(n)}.
\end{align}
As we can see, the effective cosmological constant, provided by geometric terms, is a function of the invariant ${\bf T}^{(n)}$. The limit $n=1$, where the gravitational Lagrangian is a function of the trace of the energy-momentum tensor $T$, is called ``$\Lambda(T)$ gravity'' and was studied in detail in Ref.~\cite{Poplawski:2006ey}.
\section{Linearized equations}
\label{Sec:LinearizedEq}
Let us now consider a standard scalar field Lagrangian, containing an unknown potential $V(\phi)$ and the kinetic term $\partial_\alpha \phi \, \partial^\alpha \phi$, as follows
\begin{align}
{\cal L}_{\rm m}= -\frac{1}{2} \, \partial_\alpha \phi \, \partial^\alpha \phi - V(\phi).
\end{align}
By this choice, the energy-momentum tensor becomes
\begin{align}
T_{\mu\nu} = \partial_\mu \phi \, \partial_\nu \phi -g_
{\mu\nu}\left( \frac{1}{2} \partial^\alpha\phi \, \partial_\alpha \phi+V(\phi)\right),
\end{align}
whose trace reads
\begin{align}
T^\alpha_{\phantom{\alpha}\alpha}= -\partial^\alpha \phi \, \partial_\alpha \phi -4 V(\phi).
\end{align}
Considering the definitions \eqref{Tnmunu} and \eqref{Tn}, the derivatives of ${\bf T}^{(n)}$ with respect to the scalar field $\phi$ and its derivative $\partial_\mu \phi$ read, respectively
\begin{align}
	\frac{\partial {\bf T}^{(n)}}{\partial  \phi}=& -n \, V^\prime (\phi) \, {\bf T}^{(n-1)},
	\\
	\frac{\partial {\bf T}^{(n)}}{\partial \partial_\mu \phi}=&2n \, {{\bf T}^{\mu}_{\phantom{\mu}\alpha}}^{(n-1)} \, \partial^\alpha \phi -n  {\bf T}^{(n-1)} \, \partial^\mu \phi,
\end{align}
where the prime denotes the derivative with respect to $\phi$. A Klein-Gordon-like equation occurs after varying the total action with respect to the scalar field, providing:
\begin{align}
\nabla_\mu \left[-\partial^\mu \phi \left(1+\frac{n}{2\kappa^2}  f_{{\bf T}^{(n)}} {\bf T}^{(n-1)} \right)+\frac{n}{\kappa^2} f_{{\bf T}^{(n)}} {{\bf T}^{\mu}_{\phantom{\mu}\alpha}}^{(n)} \, \partial^\alpha \phi
\right] + \left(1+\frac{n}{2\kappa^2}f_{{\bf T}^{(n)}} {\bf T}^{(n-1)}\right) V^\prime(\phi) = 0 \label{SFE}.
\end{align} 
In what follows, we consider the linearized equations of this model, with the aim to study the GW equations and point out the differences between $f(R,{\bf T}^{(n)})$ extended theory and standard GR.

We consider a general energy-momentum tensor of matter fields ${{\bf T}^{\mu}_{\phantom{\mu}\nu}}^{(n)}$, to obtain the linearized field equations of $f(R, {{\bf T}^{\mu}_{\phantom{\mu}\nu}}^{(n)})$ gravity. To this purpose, let us start with the definition of the higher-order invariant ${{\bf T}^{\mu}_{\phantom{\mu}\nu}}^{(n)}$ in terms of the energy momentum tensor, that is
\begin{align}
    {\bf T}^{(n)}=& T^{\alpha_1}_{\phantom{\alpha_1}\alpha_2} \, T^{\alpha_2}_{\phantom{\alpha_2}\alpha_3} \, \cdots \, T^{\alpha_{n}}_{\phantom{\alpha_{n}}\alpha_1}.
    \label{defTn}
\end{align}
The first order of perturbations of Eq. \eqref{defTn} yields:
\begin{align}
   \delta {\bf T}^{(n)}=& \delta T^{\alpha_1}_{\phantom{\alpha_1}\alpha_2} \, \bar{T}^{\alpha_2}_{\phantom{\alpha_2}\alpha_3} \, \cdots \, \bar{T}^{\alpha_{n}}_{\phantom{\alpha_{n}}\alpha_1}+
   \bar{T}^{\alpha_1}_{\phantom{\alpha_1}\alpha_2} \, \delta T^{\alpha_2}_{\phantom{\alpha_2}\alpha_3} \, \cdots \, \bar{T}^{\alpha_{n}}_{\phantom{\alpha_{n}}\alpha_1}+\cdots
   \\
   =&n\, \delta T^{\alpha_1}_{\phantom{\alpha_1}\alpha_2} \, \bar{T}^{\alpha_2}_{\phantom{\alpha_2}\alpha_3} \, \cdots \, \bar{T}^{\alpha_{n}}_{\phantom{\alpha_{n}}\alpha_1}
    \\
   =&n\, \delta T^{\alpha_1}_{\phantom{\alpha_1}\alpha_2} \, \bar{{\bf T}}^{(n-1) \, \alpha_2}_{\phantom{(n-1) \, \alpha_2}\alpha_1}
   \\
   =&n\, \delta T^{\mu}_{\phantom{\mu}\nu} \, \bar{{\bf T}}^{(n-1)\, \nu}_{\phantom{(n-1)\, \nu}\mu},
\end{align}
where quantities denoted by the lines on the top are evaluated at the zero-th order. The above result can be replaced into the first-order variation of the function $f$, namely
\begin{align}
     \delta f\left(R,{\bf T}^{(n)}\right) = \overline{f_R} \, \delta R + \overline{f_{{\bf T}^{(n)}}} \, \delta {\bf T}^{(n)},
\end{align}
by means of which the field equations \eqref{FE} can be computed at the first order. Hereafter, the line on the top of the given function is meant for the evaluation at the zero-th order. In order to investigate the GW modes and the linearized field equations, we consider the perturbation of the metric tensor $g_{\mu\nu}$ around a flat background metric $\eta_{\mu\nu}$ and the perturbation of the scalar field $\phi$ around its background value $\phi_0$, \emph{i.e.}
\begin{align}
g_{\mu\nu} =& \eta_{\mu\nu} + h_{\mu\nu} + {\cal O}\left(h^2 \right), \label{perturbmetric}
\\
	\phi = & \phi_0 + \delta \phi + {\cal O}\left(\delta \phi^2 \right),
\end{align}
where $\left|\eta_{\mu\nu}\right| \ll \left|h_{\mu\nu}\right|$ and $ \left|\phi_0\right| \ll \left|\delta \phi\right|$. In light of this assumption, the energy-momentum tensor of the scalar field, up to the first order, can be written as: 
\begin{align}
T^{\mu}_{\phantom{\mu}\nu}\simeq&-\delta^\mu_\nu \left(V_0+V^\prime \, \delta \phi\right) + {\cal O}\left(h^2\right),
\label{1storderTmunu}
\end{align}
with the definitions $V_0 \equiv V(\phi_0)$ and  $V'= \left.\frac{dV(\phi)}{d \phi}\right|_{\phi = \phi_0}$.  Substituting Eq. \eqref{1storderTmunu} into Eqs. \eqref{Tnmunu} and \eqref{thetamunu}, we get, respectively
\begin{align}
{{\bf T}^{\mu}_{\phantom{\mu}\nu}}^{(n)}\simeq &(-V_0)^n \delta^\mu_\nu \left(1+n \frac{V^\prime}{V_0} \, \delta \phi\right) + {\cal O}\left(h^2\right),
\label{1storderTnmunu}
\end{align}
\begin{align}
	\Theta^{(n)}_{\mu\nu} \simeq {\cal O}\left(h^2\right).
	\label{1storderthetanmunu}
\end{align}
Using Eqs. \eqref{1storderTnmunu} and \eqref{1storderthetanmunu}, the field equations can be evaluated at the zero-th and first-order of perturbations, providing respectively:
\begin{align}
\overline{f} = 2\kappa^2 V_0
\end{align}
and
\begin{align}
	 \left(R_{\mu\nu}^{(1)}-\frac{1}{2}\eta_{\mu \nu} R^{(1)}\right) \overline{f_R} +\overline{f_{RR}} \left(\eta_{\mu\nu} \square -\partial_\mu \partial_\nu \right) R^{(1)} &
	 \nonumber \\   -\left[-\kappa^2 V^\prime+2n \overline{f_{T^{(n)}}} (-V_0)^n \frac{V^\prime}{V_0} \right] \eta_{\mu\nu} \, \delta \phi +4 n \overline{f_{RT^{(n)}}} (-V_0)^n \frac{V^\prime}{V_0} \left(\eta_{\mu\nu} \square -\partial_\mu \partial_\nu \right) \, \delta \phi =&0,
	 \label{fieldEqs}
\end{align}
whose trace equation is 
\begin{align}
-\overline{f_R} R^{(1)} +3 \overline{f_{RR}} \square R^{(1)} =\left[-4\kappa^2 V^\prime+8n\overline{f_{T^{(n)}}} (-V_0)^n \frac{V^\prime}{V_0} \right]  \, \delta \phi -12 n \overline{f_{RT^{(n)}}} (-V_0)^n \frac{V^\prime}{V_0} \, \square \, \delta \phi ,
\label{traceEqs}
\end{align}
with $R_{\mu\nu}^{(1)}$ and $R^{(1)}$ being the Ricci tensor and the Ricci scalar evaluated at the first order.

By similar computations, also the Klein-Gordon equation \eqref{SFE} can be evaluated at the lowest order of perturbations. It yields:
\begin{align}
	\overline{f_{T^{(n)}}} = -\frac{2\kappa^2 V^\prime}{n(-V_0)^{n-1}},
\end{align} 
and
\begin{align}
	-\left[1+\frac{n}{\kappa^2}\overline{f_{T^{(n)}}}(-V_0)^{n-1}\right] \, \square \, \delta \phi &
	\nonumber\\+ \left[V^{\prime \prime}+\frac{2n(n-1)}{\kappa^2}\overline{f_{T^{(n)}}}(-V_0)^{n-1} \frac{V^\prime}{V_0}+\frac{8n^2}{\kappa^2}\overline{f_{T^{(n)}T^{(n)}}}(-V_0)^{n-1} \frac{V^\prime}{V_0}\right]\, \delta \phi & \nonumber \\
	+\frac{2n}{\kappa^2} \overline{f_{T^{(n)}R}} (-V_0)^{n-1}\, R^{(1)} &=0.
		\label{KGEq}
\end{align}
After few computations we end up having two d'Alembert-like equations, respectively with respect to the scalar curvature and the scalar field. We also assume $\phi_0$ as the minimum for the potential, so that we can write
\begin{align}
V=V_0 + \frac{1}{2} a (\delta \phi)^2,
\end{align} 
with $a$ being a real constant. With this choice, the field equations \eqref{fieldEqs}, the trace equation \eqref{traceEqs} and the equation for the scalar field \eqref{KGEq} can be respectively recast as
\begin{align}\label{eq:36}
\left(R_{\mu\nu}^{(1)}-\frac{1}{2}\eta_{\mu \nu} R^{(1)}\right) \overline{f_R} +\overline{f_{RR}} \left(\eta_{\mu\nu} \square -\partial_\mu \partial_\nu \right) R^{(1)}  =&0,
\\ \label{eq:47}
-\overline{f_R} R^{(1)} +3 \overline{f_{RR}}\; \square R^{(1)} =&0,
\\ \label{eq:40}
\left( \square - V^{\prime \prime}\right)\, \delta \phi
-\frac{2n}{\kappa^2} \overline{f_{T^{(n)}R}} (-V_0)^{n-1}\, R^{(1)}=&0 .
\end{align}

Up to the first order in perturbations, the explicit expression of the Ricci tensor $R_{\mu\nu}^{(1)}$ and the Ricci scalar $R^{(1)}$ (under the \emph{ansatz} \eqref{perturbmetric}) is
\begin{align}
    R^{(1)}_{\mu\nu} =& \frac{1}{2} \left( \partial_\sigma \partial_\mu h^\sigma_\nu + \partial_\sigma \partial_\nu h^\sigma_\mu - \partial_\mu \partial_\nu h - \square h_{\mu\nu} \right),
    \\
    R^{(1)} =& \partial_\mu \partial_\nu h^{\mu\nu} - \square h_{\mu\nu},
\end{align}
where $h$ is the trace of metric perturbation $h_{\mu\nu}$. Finally, by means of the definition
\begin{align}
\bar{h}_{\mu\nu} = h_{\mu\nu}-\frac{1}{2}\eta_{\mu\nu}h+ \frac{\overline{f_{RR}}}{\overline{f_R}} \eta_{\mu\nu} R^{(1)},
\end{align}
Eq.~\eqref{eq:36} becomes
\begin{align}
	\square \bar{h}_{\mu\nu}=0.
\end{align}
Moreover, by using the plane wave \emph{ansatz}, corresponding to the standard Fourier decomposition
\begin{align}
\bar{h}_{\mu\nu} = \hat{\varepsilon}_{\mu\nu}(p^\alpha)\; \exp(i p_\mu x^\mu),
\end{align}
with $p^\alpha p_\alpha = 0$, Eq.~\eqref{eq:47} can be analytically solved providing
\begin{align}
R^{(1)} = \hat{R}(q^\alpha) \;  \exp(iq_\mu x^\mu),
\end{align}
where $q^\alpha q_\alpha = -m^2 =- \overline{f_R} / (3 \overline{f_{RR}})$. 
Finally, to the purpose of finding analytic solutions to Eq.~\eqref{eq:40}, we assume the scalar field perturbation $\delta \phi$ to have the form:  
\begin{align}
\delta \phi = \hat{\phi}(s^\alpha) \; \exp(i s_\mu x^\mu),
\label{ansatzscalarfield}
\end{align}
where $s^\alpha s_\alpha = -m_\phi^2$. Due to the \emph{ansatz} \eqref{ansatzscalarfield}, Eq.~\eqref{eq:40} can be recast as
\begin{align}
\left(\square - V^{\prime \prime}\right)\, \hat{\phi}\; \exp(i s_\mu x^\mu)=&
\frac{2n}{\kappa^2} \overline{f_{T^{(n)}R}} (-V_0)^{n-1}\, \hat{R} \;  \exp(iq_\mu x^\mu).
\label{Eq40'}
\end{align}
When the function contains mixed terms, namely when $\overline{f_{T^{(n)}R}}\neq 0 $, it turns out that $s^\mu=q^\mu$, $m^2 = m^2_\phi$ and
\begin{align}
m_\phi^2= V^{\prime \prime}+ 
\frac{2n}{\kappa^2} \overline{f_{T^{(n)}R}} (-V_0)^{n-1}\, \frac{\hat{R}}{\hat{\phi}}.
\end{align}
On the other hand, in the absence of mixed terms, that is when the function $f\left(R,{\bf T}^{(n)}\right)$ can be recast as the sum $f\left(R,{\bf T}^{(n)}\right) =F_1\left(R\right) + F_2\left({\bf T}^{(n)}\right)$, Eq.~\eqref{Eq40'} reduces to
\begin{align}
\left(\square - V^{\prime \prime}\right)\, \hat{\phi}\, \exp(is_\alpha x^\alpha)=&0, 
\end{align}
with $m^2_\phi = V^{\prime \prime}$.

\section{Extra force}
\label{Sec:EF}

We show now that non-vanishing divergences of the matter energy-momentum tensor in $f(R,{\bf T}^{(n)})$-gravity result in a non-geodesic motion of  test particles. Computations are pursued by following the procedure outlined \emph{e.g.} in Refs.~\cite{Bertolami:2007gv, Harko:2010zi, Harko:2008qz, Harko:2010mv}. To this purpose, we assume matter to be described by a perfect fluid, so that the corresponding energy-momentum tensor reads:
\begin{align} 
T^{\mu\nu}=p \, g^{\mu \nu } + u^{\mu }\, u^{\nu } \,  (p + \rho),
\end{align}
where $\rho$ is the energy density, $p$ is the pressure and $u^\mu={\rm d}x^\mu / {\rm d}s$ is the four-velocity, with $u^\mu u_\mu=-1$. Considering the definition \eqref{Tnmunu}, the rank-two tensor ${\bf T}^{\mu\nu \; (n)}$ can be recast in terms of pressure and density as: 
\begin{align}
{\bf T}^{\mu\nu \; (n)}=p^n \,  g^{\mu\nu} + u^{\mu } \,u^{\nu } \, \left[p^n - \right(- \rho\left)^n \right],
\end{align}
so that its trace becomes
\begin{align}
{\bf T}^{(n)}= 3 p^n+(-\rho)^n .
\label{traceperfectfluid}
\end{align}
Introducing the projection operator, defined through the relations
\begin{eqnarray}
    && h_{\mu\nu} = g_{\mu\nu}+u_\mu u_\nu, 
    \\
    && h_{\mu\nu} u^\nu=0,
\end{eqnarray}
 and making use of the identity $u^\mu \, \nabla_\nu u_\mu=0$, the contraction of Eq.~\eqref{cons} with $h^\lambda_\nu$ yields:
\begin{align}
\kappa^2 h^\lambda_\nu  \nabla^\nu p + \kappa^2 (p+\rho) u^\mu \nabla_\mu u^\lambda =& h^\lambda_\nu \nabla^\nu \left[ f_{\bf T^{(n)}} \, \left( np^n -n {\cal L}_{\rm m} p^{n-1} +\frac{n}{2} {\cal L}_{\rm m} {\bf T^{(n-1)}} -np {\bf T^{(n-1)}}\right)  \right] 
\nonumber \\ &+ f_{\bf T^{(n)}}  u^\mu \, \nabla_\mu u^\lambda \left\{n \left[p^n-(-\rho)^n\right] - n {\rm L}_{\rm m} \left[ p^{n-1}-(-\rho)^{n-1}\right]-n {\bf T^{(n)}}(p+\rho) \right\}
\nonumber \\ 
&-\frac{1}{2} f_{\bf T^{(n)}} h^\lambda_\nu \nabla^\nu {\bf T^{(n)}},
\label{nonC}
\end{align}
where we assumed $\displaystyle \frac{\partial^2 {\cal L}_{\rm m}}{ \partial g^{\mu\nu} \, \partial g^{\alpha \beta}} = 0$. The latter \emph{ansatz} is motivated by the fact that the second derivative of ${\cal L}_{\rm m}$ with respect to the metric vanishes for most of matter Lagrangians. This is the case, for instance, of the scalar field Lagrangian ${\cal L}_{\rm m} = -\frac{1}{2} \partial_\mu \phi \, \partial^\mu \phi +V(\phi)$ or some perfect fluid Lagrangian such as ${\cal L}_{\rm m}= -\rho$ or ${\cal L}_{\rm m}= p$. 
The extra force can be explicitly computed by starting from Eq.~\eqref{nonC} and using
\begin{align}
    u^\mu \, \nabla_\mu u^\nu = \frac{{\rm d}^2x^\nu}{{\rm d} s^2}+\Gamma^\nu_{\alpha \beta} \, u^\alpha \, u^\beta, 
\end{align}
thus the equation of motion for a test particle becomes
\begin{align}
\label{nongeodesic}
 \frac{{\rm d}^2x^\nu}{{\rm d} s^2}+\Gamma^\nu_{\alpha \beta} \, u^\alpha \, u^\beta := {\cal F}^\nu.
\end{align}
In the above equation, ${\cal F}^\nu$ is the extra force, defined as:
\begin{align}
{\cal F}^\lambda = \left(g^{\lambda\nu}+u^\lambda u^\nu\right)\; \frac{\nabla_\nu \left[ f_{\bf T^{(n)}} \, {\cal A}^{(n)}  \right] 
-\frac{1}{2} f_{\bf T^{(n)}}  \nabla_\nu {\bf T^{(n)}} - \kappa^2 \, \nabla_\nu p }{\kappa^2 (p+\rho)- f_{\bf T^{(n)}}  {\cal B}^{(n)}},
\label{extraForce1}
\end{align}
where
\begin{align}
    {\cal A}^{(n)} =& \left( np^n -n {\cal L}_{\rm m} p^{n-1} +\frac{n}{2} {\cal L}_{\rm m} {\bf T^{(n-1)}} -np {\bf T^{(n-1)}}\right),
    \label{An}
    \\
    {\cal B}^{(n)} =&\left\{n \left[p^n-(-\rho)^n\right] - n {\cal L}_{\rm m} \left[ p^{n-1}-(-\rho)^{n-1}\right]-n (p+\rho)  {\bf T^{(n-1)}}\right\},
    \label{Bn}
\end{align}
and $\Gamma^\nu_{\alpha\beta}$ is the Levi-Civita connection associated to the metric $g_{\mu\nu}$.
Eq.~\eqref{nongeodesic} shows that the motion is non-geodesic and the extra force ${\cal F}^\lambda$ is perpendicular to the four-velocity $u^\lambda$, \emph{i.e.} 
\begin{align}
{\cal F}^\lambda u_\lambda =0.
\end{align}
\subsection{Equation of Motion of a Test Particle}
In this subsection we use the Newtonian limit to show that the non-geodesic motion described by Eq. \eqref{nongeodesic} yields further contributions to the acceleration of a perfect fluid in $f(R,{\bf T}^{(n)})$ gravity. The equation of motion with extra force can be obtained by the following relation
	\begin{align} \label{Eq72}
	\delta S_{\rm p} = \delta \int L_p \, {\rm d}s = \delta \int \sqrt{Q} \sqrt{-g_{\mu\nu} u^\mu u^\nu}\, {\rm d}s =0,
	\end{align}
	where $L_p = \sqrt{Q} \sqrt{g_{\mu\nu}u^\mu u^\nu}$ is the Lagrangian of the test particles.
	The equations of motion can be obtained by replacing the Lagrangian $L_p$ into the Euler-Lagrange equations $\displaystyle \frac{\rm d}{{\rm d}s} \left( \frac{\partial L_p}{\partial u^\mu} \right)-\frac{\partial L_p}{\partial x^\mu} =0 $ and read as:
	\begin{align} \label{eq:eomF}
	\frac{{\rm d}^2 x^\mu}{{\rm d}s^2} + \Gamma^\mu_{\alpha\beta} \, u^\alpha u^\beta = - (g^{\mu\nu}+u^\mu u^\nu) \, \nabla_\nu \ln \sqrt{Q} 
	\end{align} 
	When $\sqrt{Q} \rightarrow 1$, Eq. \eqref{eq:eomF} describes a geodesic motion. 
	Starting from Eq. \eqref{eq:eomF}, we use the variational principle to study the Newtonian limit.	In the limit of small velocities and weak gravitational fields, the particle Lagrangian can be approximated to
	\begin{align}
	\sqrt{Q} \, \sqrt{-g_{\mu\nu}u^\mu u^\nu } \,  {\rm d}s \approx  (1+U_{\rm E})\sqrt{1+2 \Phi - \vec{v}^2} \, {\rm d}t \approx \left(1+U_{\rm E}+\Phi - \frac{\vec{v}^2}{2}\right) \, {\rm d}t,
	\end{align}
	where $\Phi$ is the Newtonian potential, $\vec{v}$ is the velocity of the fluid and where we have considered in the weak gravitational field 
	\begin{align}
		\sqrt{Q} = 1+U_{\rm E},\qquad U_{\rm E} \ll 1.
	\end{align}
	Therefore, the variational principle in the Newtonian limit becomes
	\begin{align}
	\delta \int \left(1+U_{\rm E}+\Phi - \frac{\vec{v}^2}{2}\right) \, {\rm d}t = 0.
	\end{align}
	Consequently, the total acceleration of the system reads 
	\begin{align}
	\vec{a} = -\nabla \Phi - \nabla U_{\rm E}.
	\label{accelerationwithUE}
	\end{align}
	where the first term in the RHS is the Newtonian gravitational acceleration and the second term is the effective potential, arising as a consequence of the gravitational Lagrangian extension.
	
	We can now assume a standard equation of state of the form $p =\omega \rho$, with $\omega$ constant, and a matter Lagrangian of the form ${\cal L}_{\rm m}=-\rho$. In this way, Eqs. \eqref{traceperfectfluid}, \eqref{An} and \eqref{Bn} can be written respectively as:
	\begin{align}
	{\bf T}^{(n)}=&\left[3\omega^n+(-1)^n\right]\, \rho^n,
	\\
	{\cal A}^{(n)} =& a^{(n)} \, \rho^n,
	\\
	{\cal B}^{(n)} =& b^{(n)} \, \rho^n,
	\end{align}
	with $a^{(n)}$ and $b^{(n)}$ given by
	\begin{align}
		a^{(n)} :=& n\omega^n +n \omega^{n-1}-\frac{n}{2} \left[3\omega^{n-1}+(1)^{n-1}\right] -n \omega \left[3\omega^{n-1}+(-1)^{n-1}\right],
	\\
	b^{(n)} :=& n\left[\omega^n-(-1)^n\right] +n \left[\omega^{n-1}-(-1)^{n-1}\right]-(\omega+1)\left[3\omega^{n-1}-(-1)^{n-1}\right].
	\end{align}
	We now assume $f(R,{\bf T}^{(n)}) = R+ \lambda \, G({\bf T}^{(n)})$, where $\lambda$ is a small parameter and $G({\bf T}^{(n)})$ a general function of the invariant ${\bf T}^{(n)}$. By means of this assumption, GR can be recovered as $G$ vanishes and the function $f(R,{\bf T}^{(n)})$ can be expanded in terms of $G$ and its derivatives around the fixed energy density $\rho_0$. More precisely, recasting $G({\bf T}^{(n)})$ as $G = G(\rho)$, we obtain
	 \begin{align}
	f_{{\bf T}^{(n)}}(R, {\bf T}^{(n)}) = \lambda \, G(\rho)\approx \lambda \, G(\rho_0)+\lambda \, \left. \frac{{\rm d}G(\rho)}{{\rm d}\rho} \right|_{\rho=\rho_0} \left(\rho - \rho_0 \right),
	 \end{align}
	 so that the extra force \eqref{extraForce1} reduces to
	 \begin{align} \label{eq:extraF}
	 {\cal F}^\lambda = -\alpha \left(g^{\lambda\nu}+u^\lambda u^\nu\right)  \frac{\nabla_\nu \rho }{\rho },
	 \end{align}
	 where 
	 \begin{align}
	 \alpha :=\frac{\frac{\lambda}{2} \left. \frac{{\rm d}G(\rho)}{{\rm d}\rho} \right|_{\rho=\rho_0}+\kappa^2 \omega-\frac{\lambda a^{(n)} }{n\left[3\omega^n+(-1)^n\right]}\left. \frac{{\rm d}G(\rho)}{{\rm d}\rho} \right|_{\rho=\rho_0} }{\kappa^2 (1+\omega)-\frac{\lambda b^{(n) } }{n\left[3\omega^n+(-1)^n\right]}\left. \frac{{\rm d}G(\rho)}{{\rm d}\rho} \right|_{\rho=\rho_0}}.
	 \end{align}
	 Comparing the force~\eqref{eq:extraF} with the RHS of Eq.~\eqref{eq:eomF}, we have $\sqrt{Q}(\rho)=\left(\rho / \rho_0\right)^\alpha$, which, in the weak-field limit, becomes $ \displaystyle \sqrt{Q}(\rho) \approx 1+ \alpha \ln \left(  \frac{\rho}{\rho_0}\right)$. As a consequence, the effective potential and the total acceleration in Eq.  \eqref{accelerationwithUE} turn out to be, respectively:
	 \begin{eqnarray}
	     && 	 U_{\rm E}=\alpha \ln \frac{\rho}{\rho_0},
	     \\
	     && 	\vec{a}=-\nabla \Phi - \alpha \frac{\nabla \rho }{\rho}.
	 \end{eqnarray}
	 Therefore, the additional acceleration due to the extra force reads: 
	 \begin{align}
	 \vec{a}_{\rm E} = - \alpha \frac{\nabla \rho}{\rho}.
	 \end{align}
	 	 This additional acceleration cannot be theoretically fixed, but can be phenomenologically interpreted in different contexts, such as the threshold acceleration of MOND $a_0 = 10^{10} m/s^2$ \cite{Milgrom} or the acceleration related to the Pioneer anomaly $ a_{Pioneer} = (8.5 \pm 1.3) \times 10^{-10} m/s^2$ \cite{Bertolami:2007gv}.  In this perspective, the theory can be precisely compared with observations and experiments. 
		 
	\section{Discussion and Conclusions}
	\label{Sec:concl}
	
	In this paper, we  introduced a  general gravitational Lagrangian density depending on an arbitrary function of the Ricci curvature scalar $R$ and  the $2n$-th order invariant ${\bf T}^{(n)}$ constructed by the trace of the energy-momentum tensor. We showed that the matter energy-momentum tensor conservation, \emph{i.e.} $\nabla_\mu T^{\mu\nu}=0$, is not guaranteed due to the  ${\bf T}^{(n)}$ extra terms. 
	In this framework, we  linearized  the field equations and showed that further GW modes with respect to GR emerge. These ones are massive scalar modes.	
	
The non-conservation of the energy-momentum tensor implies  that the motion of particles is non-geodesic due to the presence of an extra-force which  is perpendicular to the four-velocity.  Specifically, from the weak field limit  of $f(R,{\bf T}^{(n)})$ gravity,  it is possible to show that the total acceleration is a sum of the acceleration generated by the standard  Newtonian field and an extra-term due to ${\bf T}^{(n)}$. This result can be achieved by considering  models which reduce to GR as soon as the terms  ${\bf T}^{(n)}$ are vanishing. As pointed out at the end of the previous section, this additional acceleration can be related to MOND theory and other anomalies in test particle motions.  This fact can constitute a reliable test bed for these theories at Solar System and Galactic level.
	
In a forthcoming paper, we will select the form of  $f(R,{\bf T}^{(n)})$ function by the presence of Noether symmetry.  The Noether approach represents a well-established  physical criterion to select consistent  theories of gravity which  allow to reduce dynamics and find analytic solutions. See, {\it e.g.} \cite{Bajardi:2021tul, Urban:2020lfk, Dialektopoulos:2018qoe, Capozziello:2008ch}.	
	
\section*{Acknowledgments}
FB and SC acknowledge the support of {\it Istituto Nazionale di Fisica Nucleare} (INFN) ({\it iniziative specifiche} GINGER, MOONLIGHT2,  and QGSKY).

\end{document}